\documentclass[prb,preprint,twocolumn,10pt]{revtex4}
\usepackage{textcomp}
\usepackage{graphicx}
\usepackage{dcolumn}
\usepackage{bm}
\usepackage{times}

\begin{document}

\preprint{J. Appl. Phys. (in press), 2006}

\title{{\em Ab initio} prediction  of half-metallic 
properties for the
ferromagnetic Heusler alloys Co$_2$MSi (M=Ti, V, Cr)}

\author{Xing-Qiu Chen}
\author{R. Podloucky}
\author{P. Rogl}
\affiliation{%
Institut f\"ur Physikalische Chemie, Universit\"at Wien,
 Sensengasse 8, A 1090, Vienna, Austria
}

\begin{abstract}
By means of density functional calculations the magnetic and electronic
properties and phase stabilities of the Heusler compounds Co$_2$MSi (with
M=Ti,V,Cr,Mn,Fe,Co,Ni) were investigated. Based on the calculated results we
predict the ferromagnetic phases of the compounds Co$_2$TiSi, Co$_2$VSi and
Co$_2$CrSi to be half-metals. Of particular interest is Co$_2$CrSi because of
its high density of majority spin states at Fermi energy in combination with a
reasonably high estimated Curie temperature of 747K.  The compounds Co$_2$TiSi
and  Co$_2$VSi are thermodynamically stable, whereas Co$_2$CrSi is a
metastable phase which might be stabilized by suitable experimental
techniques.
\end{abstract}

\maketitle

\section{Introduction}
In the pioneering work of de Groot {\textrm et. al.}~\cite{Groot} NiMnSb and
PtMnSb were predicted to be half-metal ferromagnets, for which the
majority-spin states are of metallic character whereas the minority-spin
states have a gap at Fermi energy.  A larger class of Heusler alloys has such
peculiar electronic and magnetic properties~\cite{Galanakis}, which -in
combination with large magnetic moments and  high Curie temperatures- makes
these materials attractive for the design of single-spin electron
sources~\cite{Park} and spin injectors~\cite{Hashemifar,Wolf} in the field of
magnetoelectronics and related technological applications.

Within the series of Co$_2$MSi compounds (for M=Ti,V,Cr,Mn,Fe,Co,Ni),
Co$_2$FeSi and Co$_2$MnSi were studied to some extent. For Co$_2$MnSi a Curie
temperature of T$_c$ = 985 K and a total magnetic moment  of about 5
$\mu_B$~\cite{Ishida,Brown,Raphael} were measured.  For Co$_2$FeSi, the
measured Curie temperature of T$_c$ = 1100 K is the highest for all known
Heusler alloys reflecting the large total magnetic moment of about 6
$\mu_B$~\cite{Wurmehl,Wurmehlprb}.  Recently, thin films of Co$_2$MnSi and
Co$_2$MnGe~\cite{Kaemmerer, Dong} were fabricated, for which their magnetic
properties are still under debate.  Density functional theory (DFT)
calculations for Co$_2$MnSi~\cite{Ishida,Picozzi}  confirmed its half-metal
properties and the measured large magnetic moment. A conventional DFT study on
Co$_2$FeSi~\cite{Wurmehlprb} derived that Fermi energy does not fall any more
into the gap of the minority-spin states.  Treating the d-like states of Fe as
strongly localized states, the half-metal ferromagnetic property could be
recaptured, as it is claimed by experiment.~\cite{Wurmehlprb}

In the present work, by means of a DFT approach we examined the series of
Heusler alloys Co$_2$MSi (with M=Ti,V,Cr,Mn,Fe,Co,Ni) assuming they
crystallize in the typical L2$_1$  structure. For all these compounds we
derived magnetic and electronic properties as well as thermodynamical
stabilities and elastic properties. In the seven compounds, 
only for M=Ti, V, Mn, Fe compounds with the L2$_1$ structure were synthesized.
~\cite{Bushow,Gladyshevskii,Ishida,Wurmehl,Nichlescu}  The experimentally
claimed structure of  Co$_2$CoSi (or Co$_3$Si) is of the D0$_{19}$ type at
temperatures of about 1200~\textcelsius\, whereas below 1192~\textcelsius\
the compound could not be stabilized.~\cite{Ishida2} Concerning Co$_2$CrSi and
Co$_2$NiSi, no experimental data are available in literature.  Nevertheless,
for the sake of completeness and comparison, Co$_2$CrSi, Co$_2$CoSi, and
Co$_2$NiSi with the L2$_1$ structure are included in our study.

\section{Computational details}

Our DFT calculations were performed by application of the plane wave
Vienna Ab initio Simulation Package (VASP)~\cite{vasp} in its projector
augmented wave formulation for the potentials. The exchange-correlation
potential and energy were described within the generalized gradient
approximation (GGA) of Perdew and Wang~\cite{gga91} in combination with the
parametrization of Vosko {\textrm et al.}~\cite{Vosko} for spin polarized
densities.  The cubic lattice parameters were optimized self-consistently, and
care was taken to converge the total energy in terms of basis functions and of
{\bf k} points for the Brillouin zone integration. Ferromagnetic as well as
some selected antiferromagnetic orderings 
were considered. Local properties (such as projected 
densities of states and local magnetic moments)
were determined within suitably chosen spheres centered at the atomic
positions. Finally, thermodynamic and elastic stabilities were  derived by a
procedure reported in Ref.~\onlinecite{chen1}. For Co$_2$CrSi and
Co$_2$FeSi, additional studies were made by treating the d-like states as
strongly localized states in terms of a so-called LDA+U
approach~\cite{Dudarev} with only the one effective parameter U-J, similar to
the LDA+U approach of Ref.~\onlinecite{Wurmehlprb}. Our LDA+U calculations
were done on the basis of the GGA potentials as used for the conventional
calculations.

\section{results and discussions}

\begin{figure}
\includegraphics[scale=0.43]{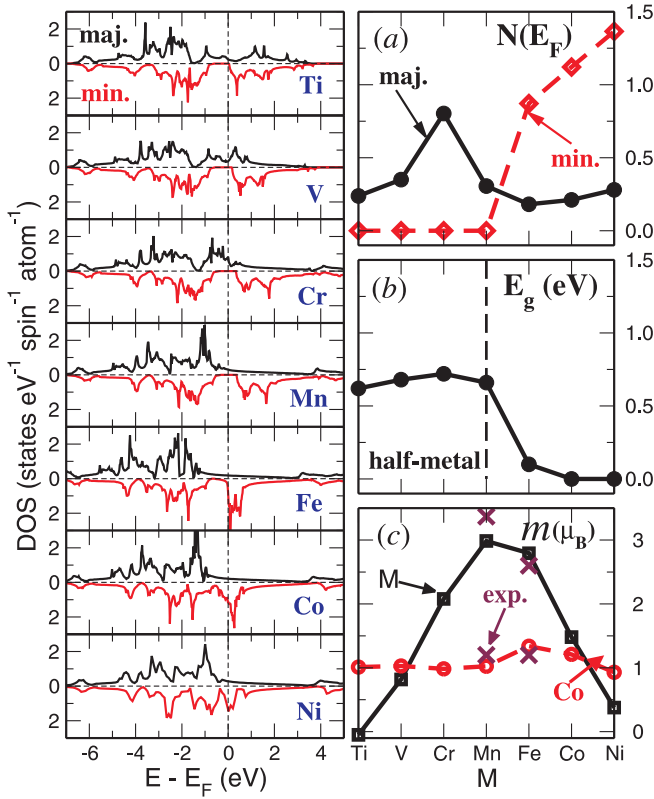}
\caption{(Color online) 
Magnetic and electronic properties derived from VASP
calculations with GGA potentials for the compounds Co$_2$MSi
(M=Ti,V,Cr,Mn,Fe,Co,Ni) with L2$_1$ crystal structure.  Panels on the left
side: ferromagnetic spin-polarized density of states split into majority-spin
and  minority-spin states.  Panels on the right side: panel $a$) shows
N(E$_\textrm{F}$), the density of states at the Fermi level for majority-spin
states (full line) and minority-spin states (broken line); panel $b$) shows
E$_\textrm{g}$ (in eV), the characteristic band gap at E$_\textrm{F}$ of the
minority-spin states; panel $c$) illustrates the trend of $m$ (in $\mu_B$),
the local magnetic moments of Co (broken line) and M atoms (full line),
together with experimental data (crosses). \label{fig1} }
\end{figure}

Fig.\ref{fig1} summarizes trends of the density of states (DOS) and
related quantities which were derived from the described VASP calculations
applying GGA potentials.  For M=Ti,V,Cr,Mn the Fermi energy E$_\textrm{F}$
lies in the pronounced gap of the minority-spin states (left side panels),
determining the half-metal character of these compounds. According to
Fig.\ref{fig1}b the width of the band gap has values in the range of 0.6 to
0.7~eV . The Heusler alloy Co$_2$FeSi, however,  is not a half-metal anymore,
because E$_\textrm{F}$ falls into an uprising peak of the minority-spin
states.  The -now indirect- band gap is only 0.1 eV wide, being strongly
reduced due to a flat tail of the mentioned peak (see corresponding left side
panel).  For the half-metal compounds, however, this pronounced peak is
totally unoccupied. For M=Co,Ni the gap vanishes, because all the d-bands are
nearly filled and the separation of the Co-d and M-d band center is now too
small for the formation of a gap. The crucial gap for the half-metal
compounds, as discussed by Galanakis {\rm et al.} for Co$_2$MnSi
\cite{Galanakis}, is due to a strong hybridization between Co-d and M-d
states, combined with large local magnetic moments and a sizeable separation
of the d-like band centers. We found that a related strong hybridization
feature (a small gap for M=Ti,V or a deep valley for M=Cr,Mn,Fe) occurs
already in the non-spin polarized DOS, for which  E$_\textrm{F}$ cuts through
a pronounced peak indicating instabiliy.  When allowing for ferromagnetic spin
polarization the Fermi energy is then pinned in the gap of the minority-spin
DOS as long as there are not too many states to be filled and the gap is
sufficiently large, which is the case for the compounds with M=Ti,V,Cr,Mn. In
these cases, the local magnetic moments $m_{\textrm{M}}$ of the M-atoms
increase linearly, whereas $m_{\textrm{Co}}$ remains rather constant (see
Fig.\ref{fig1}c).  It should be noted, that the numerical values of the local
moments depend on the choice of the localization criterion (e.g. on the choice
of radii of atomic spheres in our case), and therefore -in general- they are
not integer numbers. However, the total magnetic moments  $m_{tot}$ are
exactly integer numbers for the true half-metal compounds, as shown by
Table~\ref{tab1}.

\begin{table}[it]
\begin{center}
\caption{Results of VASP calculations with GGA potentials: magnetic moments
(total moment $m_{tot}$ in $\mu_B$/f.u., local moment $m_{\textrm{Co}}$ and
$m_{\textrm{M}}$ in $\mu_B$), density of states at Fermi level
(N($\uparrow$,E$_\textrm{F}$) for majority spin and
N($\downarrow$,E$_\textrm{F}$)  for minority spin states, in states eV$^{-1}$
spin$^{-1}$ atom$^{-1}$ ), band gaps E$_\textrm{g}$ (in eV) of the minority
spin state, and estimated Curie temperatures $T_c$ (in K)  for Co$_2$MSi
(M=Ti,V,Cr,Mn,Fe,Co,Ni). 
\label{tab1}}
\begin{ruledtabular}
\begin{tabular}{lcccccccccccccccccccccc}
 & Co$_2$TiSi & Co$_2$VSi
& Co$_2$CrSi & Co$_2$MnSi
& Co$_2$FeSi \\
\hline
$m_{\textrm{Co}}$    & 1.01 & 1.03 & 0.98 & 1.02 &  1.34   \\
$m_{\textrm{M}}$     & -0.05& 0.82 & 2.08 & 2.99 &  2.79   \\
$m_{\textrm{tot}}$     & 2  & 3  & 4  & 5  & 5.48          \\
N$_{(\uparrow,E_F)}$   & 0.24 & 0.35 & 0.80 & 0.31 & 0.18  \\
N$_{(\downarrow,E_F)}$ & 0    &0     & 0    & 0    &  0.85 \\
E$_g$                  & 0.62 & 0.68 & 0.72 & 0.66 & 0.10  \\
T$_c$                  & 385  & 566  & 747  & 928  & 1109  \\
\end{tabular}
\end{ruledtabular}
\end{center}
\end{table}

\begin{table}[it]
\begin{center}
\caption{
Calculated bulk moduli and
elastic constants ($c_{11}$, $c_{12}$, $c_{44}$ in GPa) for
the Co$_2$MSi (M= Ti, V, Cr, Mn, Fe).
\label{tab2}}
\begin{ruledtabular}
\begin{tabular}{lccccccccc}
M & Ti & V & Cr & Mn & Fe \\
\hline
B          & 215 & 216 & 227 & 221 & 204 \\
$c_{11}$   & 303 & 255 & 297 & 316 & 247 \\
$c_{12}$   & 172 & 197 & 193 & 174 & 182 \\
$c_{44}$   & 126 & 130 & 145 & 143 & 133 \\
$c^\prime$ & 65  & 29  & 52  & 71  & 33  \\
\end{tabular}
\end{ruledtabular}
\end{center}
\end{table}

\begin{table}[it]
\begin{center}
\caption{
Calculated lattice parameter (a in \AA\,) and 
enthalpies of formation ($\Delta$H, kJ (mol of atoms)$^{-1}$) 
compared to available experimental data
for Co$_2$MSi (M= Ti, V, Cr, Mn, Fe).
NM: non magnetic calculations, FM: ferromagnetic calculations.
\label{tab3}}
\begin{ruledtabular}
\begin{tabular}{lccccccccc}
\multicolumn{2}{c}{M} & Ti & V & Cr & Mn & Fe \\
\hline
a         & NM     & 5.7205 & 5.6393 & 5.5897 & 5.5582 & 5.5410 \\
          & FM     & 5.7609 & 5.6621 & 5.6295 & 5.6457 & 5.6231 \\
          & exp.   & 5.770 \footnotemark[1] & 5.659 \footnotemark[2]  
&        & 5.654 \footnotemark[3]  & 5.640 \footnotemark[4]\\
$\Delta$H & NM     & -61.5  & -32.2  & -11.7  & -9.42  & -4.7  \\
          & FM     & -64.4  & -40.7  & -29.7  & -44.9  & -33.9  \\ 
\end{tabular}
\footnotetext[1]{Ref.\onlinecite{Gladyshevskii}}
\footnotetext[2]{Ref.\onlinecite{Bushow,Nichlescu}}
\footnotetext[3]{Ref.\onlinecite{Brown}}
\footnotetext[4]{Ref.\onlinecite{Wurmehlprb}}
\end{ruledtabular}
\end{center}
\end{table}

Comparing  Co$_2$MnSi to Co$_2$FeSi, the local moment $m_{\textrm{Fe}}$ is now
smaller than $m_{\textrm{Mn}}$, and the linear trend for $m_{\textrm{M}}$ (for
M=Ti,V,Cr,Mn) is now destroyed, as shown by Fig.\ref{fig1}c . The reason is
that according to a conventional DFT GGA calculation Co$_2$FeSi is not a half
metal. This is consistent with a recent  DFT study by Wurmehl \emph{et.
al.}~\cite{Wurmehl,Wurmehlprb}, who also found that the Fermi level is not in
the band gap of the minority-spin DOS when applying GGA or the local density
approximation (LDA) for the exchange-correlation potential. Treating the
d-states as strongly localized states by means of an LDA+U approach and
choosing suitable parameters for U-J~\cite{Wurmehl}, the compound Co$_2$FeSi
became a half-metal with a local moment of 6~$\mu_B$ in agreement to
experiment. We reproduced the LDA+U results which can also be obtained from
conventional GGA calculations for a volume which is expanded by 5\% compared
to the calculated equilibrium value. It should also be mentioned that in the
case of Co$_2$MnSi, our results agree well with earlier DFT
studies.\cite{Ishida,Picozzi}

\begin{figure}
 \includegraphics[width=3in,angle=0]{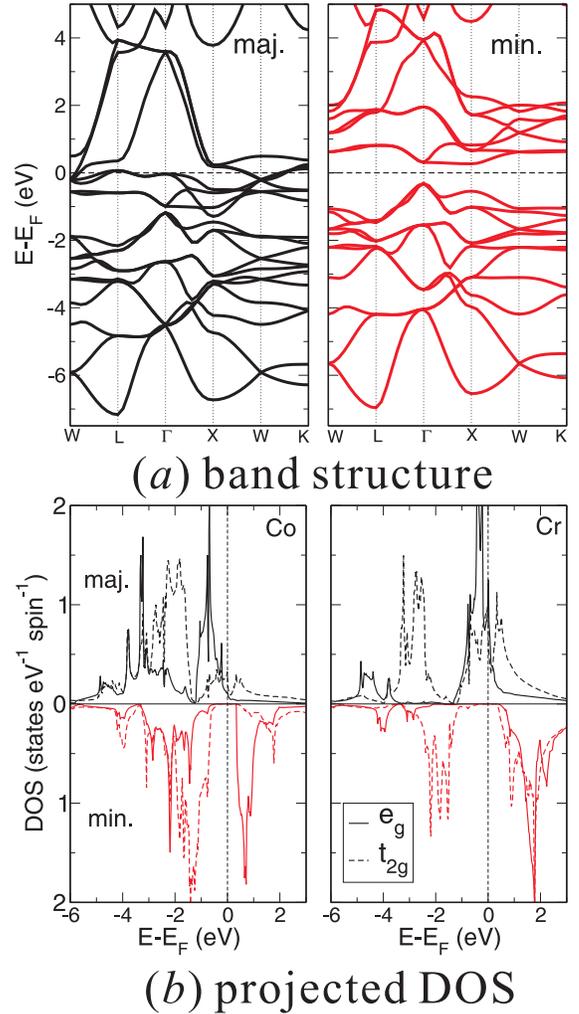}\\
 \caption{(Color online) VASP results with GGA potentials for ferromagnetic
Co$_2$CrSi: spin polarized band structure (upper panel) and projected density
of states (lower panel).
\label{fig2}}
\end{figure}

According to our results, amongst the three predicted half metals
Co$_2$MSi (M=Ti,V,Cr) the Cr compound is the most interesting one, because of
its large density of states of the majority spins at Fermi energy of
N($\uparrow$,E$_\textrm{F}$) = 0.80 states eV$^{-1}$ spin$^{-1}$ atom$^{-1}$.
This is the largest value for all known ferromagnetic
half-metals. The reason for this large value is that the Fermi energy cuts
through strongly localized states of mostly Cr-d like character, as
illustrated by the band structure and d-like DOS in Fig.~\ref{fig2}. The
contribution of Co d-states to N($\uparrow$,E$_\textrm{F}$) is very small,
because E$_F$ falls into a deep minimum of e$_g$ as well as t$_2g$ states.
According to this finding, Co$_2$CrSi is a half-metal which offers 100\%
spin-selectivity combined with a pronounced site selectivity of strongly
localized Cr states at E$_F$. This is in contrast to the well-known
half-metals  Co$_2$MnSi and Co$_2$FeSi, for which the states at E$_F$ are of
more delocalized character.

The Curie temperatures $T_{c}$ for Co$_2$MSi (M=Ti,V,Cr) were estimated
similar to a model presented in Ref.~\onlinecite{Wurmehlprb} by applying the
relation \begin{math} T_c = 23 + 181m_{tot} \end{math}, which is linear in the
total magnetic moment $m_{tot}$ per unit cell. The results are 385, 566, and
747 K for Co$_2$TiSi, Co$_2$VSi, and Co$_2$CrSi, respectively. Again, the Cr
compound is of special interest, now because of its rather large $T_{c}$,
making it interesting for technological applications.

Because the Cr d-states at E$_F$ are strongly localized one might argue that
these states have to be treated in a more suitable way e.g. in terms of the
LDA+U approach as described previously for the study of Co$_2$FeSi. In our
case, we applied the approach of Dudarev {\em et al.} ~\cite{Dudarev} which
needs only the difference $\Delta=U-J$ of the on-site Coulomb and exchange
parameters.  Similar to the study Wurmehl {\em et al.}~\cite{Wurmehlprb} we
chose $\Delta=4.8$~eV for Co, whereas for Cr $\Delta$ was varied between 0 and
3~eV. (The choice of $\Delta=0$ corresponds to the conventional GGA
calculation.) For $\Delta=2$~eV, Co$_2$CrSi is still a half-metal with a
strongly increased gap for the minority spin states of 1.9~eV  
However, for $\Delta=3$~eV the Fermi energy cuts through minority 
spin states, and the half-metal character is destroyed. The main effects 
of the localization enforced by the LDA+U treatment are an increased gap 
width and the occupation
of more majority spin states than for the conventional GGA or LDA approaches
which lowers the Fermi energy. This lowering effect could be so strong (i.e.
for larger $\Delta$), that finally E$_F$ cuts through minority spin states
below the gap. 
The applied LDA+U approach has the advantage, that only one unknown
parameter needs to be introduced, namely the difference of U minus J. Most of the
other LDA+U approaches (e.g. Ref~\onlinecite{Liechtenstein}) rely on the two
independent parameters U and J. A suitable choice of these parameters is
-however- at present not possible, because no experimental information is yet
available. For  Co$_2$CrSi being a half-metal or not,
more elaborate calculations might be necessary to determine either
the unknown parameters, or to go beyond the limitations of the LDA+U
approaches.

Measurements of the magnetic and electronic properties of
Co$_2$CrSi would give the answer. However, the preparation of suitable
samples is not straightforward, because  Co$_2$CrSi is thermodynamically
metastable. (In contrast to  Co$_2$TiSi and Co$_2$VSi, which are even known to
exist thermodynamically.) According to our {\em ab initio} calculations of the
enthalpy of formation, the compound Co$_2$CrSi is unstable against a
thermodynamical  separation into the phases Co$_2$Si, Cr and Si by  3~kJ (mol
of atoms)$^{-1}$. On the other hand, Co$_2$CrSi  is stable against
decompositions into Co$_3$Si, Cr and Si  (by 23~kJ (mol of atoms)$^{-1}$),
CoSi, Cr and Si (by 6kJ), and Co, Cr, Si (by 29.7 kJ). 
Obviously, Co$_2$CrSi might be stabilized by
advanced experimental techniques for synthesizing metastable states.
Elastically, however, the Cr compound is very stable as demonstrated by the
calculated elastic constants of c$_{11}$=297~GPa, c$_{12}$=193~GPa, and
c$_{44}$=145~GPa, and the bulk modulus of B=227~GPa (see Table \ref{tab2}).

The calculated lattice parameters listed in Table \ref{tab3}
for the known Ti-, V-, Mn-, and Fe-based half-metals are 
-as expected- in good agreement to experiment.
\cite{Gladyshevskii,Bushow,Nichlescu,Brown,Wurmehlprb}   
In this table, enthalpies of formations are also given. In particular, non
spinpolarized (NM) calcualtions are compared to ferromagnetic (FM)
calculations, which result in significant energy gains, in particular
for the Cr-, Fe-, and Mn-based alloys.
The calculated elastic constants are given by Table \ref{tab2},  which show no
anomolous behaviour. In particular, the Co$_2$CrSi alloy fits in the trend of
the other compounds, and is certainly elastically stable.
(For  Co$_2$CoSi and Co$_2$NiSi with their assumed artificial structures
the derived shear moduli $c^\prime$ are negative, indicating elastic instability).   

\section{Conclusion}
Summarizing, based on density functional calculations we predict
three ferromagnetic half-metals, namely  Co$_2$TiSi,
Co$_2$VSi, and Co$_2$CrSi.  The Cr compound is of particular
interest because of its high density of states of 100\% polarized
states at Fermi energy together with a strong selectivity of Cr-d
states and an elevated Curie temperature. We hope that our work
stimulates experiments on Co$_2$CrSi once this metastable 
phase could be synthesized.

Acknowledgements: Work supported by the Austrian Science Foundation FWF under
project nr. P16957. Parts of the calculations were made on the Schr\"odinger PC
cluster of the University of Vienna.

\end{document}